\documentclass[epsfig,12pt]{article}
\usepackage{epsfig}
\usepackage{graphicx}
\usepackage{hyperref}

\usepackage{array}
\usepackage{amsmath}
\usepackage{amssymb}

\newcommand{\beq}{\begin{equation}}   
\newcommand{\eeq}{\end{equation}}
\newcommand{\beqn}{\begin{eqnarray}}   
\newcommand{\eeqn}{\end{eqnarray}}

\begin{document}
\unitlength = 1mm

\def\de{\partial}
\def\Tr{ \hbox{\rm Tr}}
\def\const{\hbox {\rm const.}}  
\def\o{\over}
\def\im{\hbox{\rm Im}}
\def\re{\hbox{\rm Re}}
\def\bra{\langle}\def\ket{\rangle}
\def\Arg{\hbox {\rm Arg}}
\def\Re{\hbox {\rm Re}}
\def\Im{\hbox {\rm Im}}
\def\diag{\hbox{\rm diag}}


\def\QATOPD#1#2#3#4{{#3 \atopwithdelims#1#2 #4}}
\def\stackunder#1#2{\mathrel{\mathop{#2}\limits_{#1}}}
\def\stackreb#1#2{\mathrel{\mathop{#2}\limits_{#1}}}
\def\Tr{{\rm Tr}}
\def\res{{\rm res}}
\def\Bf#1{\mbox{\boldmath $#1$}}
\def\balpha{{\Bf\alpha}}
\def\bbeta{{\Bf\beta}}
\def\bgamma{{\Bf\gamma}}
\def\bnu{{\Bf\nu}}
\def\bmu{{\Bf\mu}}
\def\bphi{{\Bf\phi}}
\def\bPhi{{\Bf\Phi}}
\def\bomega{{\Bf\omega}}
\def\blambda{{\Bf\lambda}}
\def\brho{{\Bf\rho}}
\def\bsigma{{\bfit\sigma}}
\def\bxi{{\Bf\xi}}
\def\bbeta{{\Bf\eta}}
\def\d{\partial}
\def\der#1#2{\frac{\d{#1}}{\d{#2}}}
\def\Im{{\rm Im}}
\def\Re{{\rm Re}}
\def\rank{{\rm rank}}
\def\diag{{\rm diag}}
\def\2{{1\over 2}}
\def\ntwo{${\mathcal N}=2\;$}
\def\nfour{${\mathcal N}=4\;$}
\def\none{${\mathcal N}=1\;$}
\def\ntwot{${\mathcal N}=(2,2)\;$}
\def\ntwoo{${\mathcal N}=(0,2)\;$}
\def\x{\stackrel{\otimes}{,}}

\newcommand{\cpn}{CP$(N-1)\;$}
\newcommand{\wcpn}{wCP$_{N,\widetilde{N}}(N_f-1)\;$}
\newcommand{\wcpd}{wCP$_{\widetilde{N},N}(N_f-1)\;$}
\newcommand{\wcpN}{$\mathbb{WCP}(N,N)\;$}
\newcommand{\wcpK}{$\mathbb{WCP}(K,K)\;$}
\newcommand{\wcpt}{$\mathbb{WCP}(2,2)\;$}
\newcommand{\wcpf}{$\mathbb{WCP}(4,4)\;$}
\newcommand{\wcpo}{$\mathbb{WCP}(1,1)\;$}
\newcommand{\wcp}{$\mathbb{WCP}(N,\tilde N)\;$}
\newcommand{\vp}{\varphi}
\newcommand{\pt}{\partial}
\newcommand{\tN}{\widetilde{N}}
\newcommand{\ve}{\varepsilon}
\renewcommand{\theequation}{\thesection.\arabic{equation}}

\newcommand{\sun}{SU$(N)\;$}

\setcounter{footnote}0

\vfill

\begin{titlepage}

\begin{flushright}
\end{flushright}

\begin{center}
{  \Large \bf  
 Flowing Between String Vacua for the   
\\[1mm]
Critical Non-Abelian Vortex with
\\[2mm]
 Deformation of \ntwo Liouville theory
 }

\vspace{5mm}

{\large  \bf A.~Yung$^{\,\,a,\,b}$}
\end {center}

\begin{center}

$^{a}${\it National Research Center ``Kurchatov Institute'', 
Petersburg Nuclear Physics Institute, Gatchina, St. Petersburg
188300, Russia}\\
$^{b}$ {\it Higher School of Economics, National Research University, St. Petersburg 194100, Russia}\\

\end{center}

\vspace{5mm}

\begin{center}
{\large\bf Abstract}
\end{center}

It has been  shown that  non-Abelian solitonic vortex string supported in four-dimensional (4D)
\ntwo supersymmetric QCD (SQCD) with the U(2) gauge group
and $N_f=4$  quark flavors  becomes a critical superstring. This string propagates in the ten-dimensional space formed by a product of the flat 4D space and an internal space given by a Calabi-Yau noncompact threefold, namely, the conifold.
The spectrum of low lying closed string states  was found and interpreted
as a spectrum of hadrons in 4D \ntwo  SQCD. In particular,
the lowest string state appears to be a massless BPS baryon associated with the deformation of the complex structure modulus
$b$ of the conifold. It was recently shown that the Coulomb branch of the associated string sigma model
which opens up at strong coupling can be described by \ntwo Liouville theory. Building on these results we switch on quark masses in  4D \ntwo SQCD and study the interpolation of the initial U(2) SQCD with $N_f=4$ quarks to the final SQCD with the 
 U(4) gauge group
and $N_f=8$  quarks. To find the  true string vacuum which arises due to the mass deformation we solve the effective supergravity equations of motion associated with the deformed world sheet Liouville theory.  We show that the massless BPS baryon $b$ survives the deformation and that finding of the  spectrum of low lying massive hadrons in the final SQCD is linked to the Calogero problem.

\vspace{2cm}

\end{titlepage}

\newpage


\newpage

\section {Introduction } 
\label{intro}
\setcounter{equation}{0}

It was shown in \cite{SYcstring}  that the non-Abelian solitonic vortex string  in four-dimensional 
(4D) \ntwo  supersymmetric QCD (SQCD)  with the U($N=2$) gauge group and $N_f = 2N=4$
flavors of quark hypermultiplets behaves as a critical superstring. Non-Abelian vortices were first found in 
\ntwo  SQCD with the gauge group U$(N)$ and $N_f \ge N$ flavors of quarks
\cite{HT1,ABEKY,SYmon,HT2}. The non-Abelian vortex string is 1/2
BPS saturated and, therefore,  has \ntwot supersymmetry on its world sheet.
In addition to four translational moduli  the non-Abelian string carries orientational  moduli, as well as the size moduli if $N_f>N$
\cite{HT1,ABEKY,SYmon,HT2} (see \cite{Trev,Jrev,SYrev,Trev2} for reviews). Their dynamics
are described by the effective two-dimensional (2D) sigma model on the string world sheet, the so-called \ntwot supersymmetric weighted CP  model ($\mathbb{WCP}(N,N_f-N)$). 

 For $N_f=2N$
the world sheet sigma model becomes conformal. Moreover, for $N=2$ the 
number of the orientational/size moduli  is six and they can be combined with 
four translational moduli to form a ten-dimensional  space required for a
superstring to become critical \cite{SYcstring,KSYconifold}. In this case the target space of the world sheet  theory on the non-Abelian vortex string is
 $\mathbb{R}^4\times Y_6$, where $Y_6$ is a noncompact six dimensional Calabi-Yau (CY) manifold, the conifold \cite{Candel}, see \cite{NVafa} for a review. Moreover, the theory of the critical vortex
string at hand was identified as the superstring theory of type IIA \cite{KSYconifold}.
The
spectrum of low lying closed string excitation was found in  \cite{KSYconifold,SYlittles}.

A version of the string-gauge duality for 4D SQCD was proposed \cite{SYcstring}: at
weak coupling this theory is in the Higgs phase and can be described in terms
of quarks and Higgsed gauge bosons, while at strong coupling hadrons of this theory can be understood as closed string states formed by the non-Abelian vortex string. We  call this approach a ''solitonic string-gauge duality''.

Most of massless  and massive  string modes have  non-normalizable wave functions over the conifold $Y_6$, i.e. they are not localized in 4D 
and cannot be interpreted as dynamical states in 4D theory, in particular there are no massless 4D gravitons in the physical spectrum  \cite{KSYconifold}.
However, an excitation associated with the deformation of the complex structure modulus $b$ of $Y_6$ has (logarithmically) normalizable wave function   and was  interpreted  as a massless baryon in the spectrum of hadrons of 4D \ntwo SQCD  \cite{KSYconifold}.

To analyze the massive states, a different approach was chosen
similar to the one, used for Little String Theories  ( see  \cite{Kutasov} for a review). It is  based
on the  equivalence \cite{GVafa} between the 
critical string on the conifold and non-critical $c=1$ string containing the Liouville 
field and a compact scalar  at 
the self-dual radius (united into a complex scalar of \ntwo Liouville theory \cite{Ivanov,KutSeib})~\footnote{In \cite{GVafa} this equivalence was shown for topological versions of the string theories.}.  
Later similar correspondence was proposed (and treated as a holographic AdS/CFT-type duality)
for the critical string on certain  non-compact CY spaces with isolated singularity in so-called double scaling limit, and non-critical $c=1$ string with an additional Ginzburg-Landau
\ntwo superconformal system  \cite{GivKut,GivKutP}. In the conifold case, this extra Ginzburg-Landau
conformal field theory (CFT) is absent.

Recently this equivalence was demonstrated in a more direct way. Namely, it was shown in \cite{GIMMY}
 that  Coulomb branches of  world sheet $\mathbb{WCP}(N,N)$ models  on non-compact toric CY manifolds with an isolated singularity which open up at strong coupling can be
 described by \ntwo Liouville theory with background charge depending on $N$. This was shown first in the
   large $N$ approximation and then extrapolated  to an exact equivalence.

The above equivalence was used in \cite{SYlittles,SYlittmult} to find the low lying spectrum of hadrons in
4D \ntwo SQCD with gauge group U(2) and $N_f=4$ quark flavors.

Although the solitonic string-gauge duality program looks rather generic, one of its main limitations is that so far only one example of 4D gauge theory supporting non-Abelian strings which can be quantized using well developed string theory methods was found, namely \ntwo SQCD with gauge group U(2) and $N_f=4$ quark flavors. The reason is that the string sigma model ($\mathbb{WCP}(N,N)$) living on the world sheet of non-Abelian string is conformal for $N_f=2N$ and its target space has critical dimension 
equal to ten for $N=2$ \cite{SYcstring}.

In this paper we make a step towards broadening of the class of 4D \ntwo SQCDs where the solitonic string-gauge duality can be applied, see also \cite{Y_NSflux,Y_NSflux22}. To this end we introduce quark masses in \ntwo SQCD and changing values of mass parameters interpolate between SQCDs with different gauge groups and numbers of quark flavors. 

Quark masses in 4D SQCD induce so-called twisted masses of fields in the world sheet $\mathbb{WCP}(N,N)$ model breaking its conformal
invariance. We repeat the derivation in \cite{GIMMY} and reduce the Coulomb branch  of the mass-deformed $\mathbb{WCP}(N,N)$ model
to the  deformed \ntwo Liouville  theory, where the mass deformation boils down to a non-trivial  metric of the target space. However this '' classical'' metric still cannot be used for the string quantization since the conformal invariance of the model is broken by the mass deformation.

To find a true string vacuum we solve effective supergravity equations using the  classical  metric of the mass-deformed \ntwo Liouville theory 
 only as  initial conditions  for  the true metric at large values of the Liouville field $\phi$, where the mass deformation is small. 

Solving the gravity equations of motion and finding the true quantum metric and the dilaton for the mass-deformed \ntwo Liouville theory allows us to interpolate between two 4D SQCDs. Namely, starting from \ntwo SQCD with gauge group U(2) and $N_f=4$ which supports
critical non-Abelian vortex string and reducing the mass parameter  ''integrating extra quarks in'' we interpolate to \ntwo SQCD with 
gauge group U(4) and $N_f=8$ quark flavors. We show that the $b$-baryon survives the deformation and remains massless in the final SQCD. Instead massive states ''feel'' the naked singularity which is present in the metric and finding of their spectrum is
linked to the  Calogero problem with the ''falling to the center'' $1/\phi^2$-type potential associated with the singularity. 

The paper is organized as follows. In Sec. \ref{sec:NAstring} we review 4D \ntwo SQCD which supports non-Abelian strings and 
\wcpN models  arising as world sheet theories  on these strings focusing on  conformal cases $N_f=2N$. Next, we  describe  our mass deformation which interpolates from  SQCD with U(2) gauge group and  $N_f=4$  to SQCD with U(4) gauge
group and $N_f=8$. We also review the massless $b$-baryon associated with the complex structure modulus of the conifold in the former theory. In Sec. \ref{wcp=liouville} we review the derivation of  \ntwo Liouville theory from the \wcpN world sheet model
at strong coupling and then describe its mass deformation. In Sec. \ref{sec:gravity} we study effective supergravity equations of motion associated with the mass-deformed Liouville world sheet model and find their solution which describes the true string vacuum.
In Sec. \ref{sec:spectrum} we continue using the  gravity approach and consider the tachyon equation of motion for the vertex operators of the mass-deformed theory. We show that the $b$-baryon remains massless in the mass-deformed theory and discuss qualitatively  the structure of the  spectrum of massive string states pointing out that finding of this spectrum is related to the Calogero problem. Sec. \ref{sec:conclusion} contains our Conclusions.

\section {Non-Abelian  vortex string}
\label{sec:NAstring}
\setcounter{equation}{0}

\subsection{Four-dimensional \boldmath{${\mathcal N}=2\;$} 
 SQCD}
\label{sec:SQCD}

As was already mentioned, non-Abelian vortex strings were first found in 4D
\ntwo SQCD with the gauge group U$(N)$ and $N_f \ge N$ quark flavors 
supplemented by the Fayet-Iliopoulos (FI)  term  \cite{FI} with parameter $\xi$
\cite{HT1,ABEKY,SYmon,HT2}, see for example, \cite{SYrev} for a detailed review of this theory.
The field content is as follows. The \ntwo vector multiplet
consists of the  U(1)
gauge field and the SU$(N)$  gauge field which can be combined in the matrix $(A_{\mu})^k_l$ 
  plus
complex scalar fields $a^k_l$,  and their Weyl fermion superpartners.
The $N_f$ quark multiplets of  the U$(N)$ theory consist
of   the complex scalar fields
$q^{kA}$ and $\tilde{q}_{Ak}$ (squarks) and
their   fermion superpartners, all in the fundamental representation of 
the SU$(N)$ gauge group.
Here  $\mu=1,...,4$ is the 4D Minkowski index, $k,l=1,..., N$ are  color indices,
while $A$ is the flavor index, $A=1,..., N_f$. Below we briefly describe how we can interpolate between SQCDs with different gauge groups and number of quark flavors considering different limits of quark masses.

At weak coupling $g^2\ll 1$, this theory is in the Higgs phase in which adjoint scalars develop vacuum expectation values (VEVs)
\beq
\langle a \rangle = - \frac1{\sqrt{2}}
 \left(
\begin{array}{ccc}
m_1 & \ldots & 0 \\
\ldots & \ldots & \ldots\\
0 & \ldots & m_N\\
\end{array}
\right),
\label{avev}
\eeq
 where we select a vacuum where the first $N$ quark flavors are massless at zero  FI parameter $\xi$, while $m_A$ are quark masses. 
Adjoint VEVs break U$(N)$ gauge group down to U(1)$^N$, masses of off-diagonal gauge bosons are given by 
$M^k_l \sim |m_k-m_l|$. If certain quark masses coincide adjoint VEVs leave  certain non-Abelian subgroups of U$(N)$ unbroken, see below. Masses of $q^{kA}$ and $\tilde{q}_{Ak}$ quarks are equal to $|m_k-m_A|$.

At non zero $\xi$ following diagonal components of squarks also develop VEVs
\beq
\langle q^{kk}\rangle =\sqrt{
\xi}\, \quad {\rm no\, summation}, \qquad k=1,...,N 
\label{qvev}
\eeq
with all other components equal to zero.

 These VEVs break 
the U$(N)$ gauge group
Higgsing  all gauge bosons. The Higgsed gauge bosons combine with the screened quarks to form long \ntwo multiplets with mass $m_G \sim g\sqrt{\xi}$ in the limit of zero quark masses.

In this limit the global flavor SU$(N_f)$ is also broken down by quark VEVs to the so-called color-flavor
locked group. The resulting global symmetry is
\beq
 {\rm SU}(N)_{C+F}\times {\rm SU}(N_f-N)\times {\rm U}(1)_B,
\label{c+f}
\eeq
see \cite{SYrev} for more details. 

The unbroken global U(1)$_B$ factor above is identified with a baryonic symmetry. Note that 
what is usually identified as the baryonic U(1) charge is a part of  our 4D theory  gauge group.
 ``Our" U(1)$_B$
is  an unbroken by squark VEVs combination of two U(1) symmetries;  the first is a subgroup of the flavor 
SU$(N_f)$, and the second is the global U(1) subgroup of U$(N)$ gauge symmetry.

In this paper we consider SQCDs with $N_f=2N$. In these cases coupling constants in both the 4D SQCD and the world sheet 
$\mathbb{WCP}(N,N)$ model on the non-Abelian string does not run. However, the conformal
invariance of the 4D theory is explicitly broken by the
FI parameter $\xi$, which defines the VEVs of quarks. The FI
parameter is not renormalized. We also assume that $N$ is even, $N=2K$, where $K$ is integer.

Below we consider a special choice of quark masses,
\beq
m_{A+N} =m_A, \qquad A=1,...,N
\label{extra_masses}
\eeq
which ensures that ''extra'' quarks  with $A=(N+1),...,2N$ have the same masses as the  first $N$ ones. We also assume
that 
\beq
m_A =0, \quad {\rm for} \;\;\; A=1,...,K; \qquad m_A =M, \quad {\rm for}\;\;\; A=(K+1),...,N,
\label{masses}
\eeq
so we have only one quark mass parameter $M$ to interpolate between different SQCDs.

Consider first the limit $M\to\infty$. In this limit gauge fields $(A_{\mu})^k_{k'}$ and $(A_{\mu})^l_{l'}$ are massless (at $\xi=0$),
while $(A_{\mu})^k_{l}$ and $(A_{\mu})^l_{k}$ together with their superpartners become infinitely heavy and decouple, $k,k'=1,...,K$,  
$l,l'=(K+1),...,N$. Therefore, the gauge group U$(N)$ is broken down to U$(K)\times{\rm U}(K)$.  Similarly quarks $q^{kB}$ and 
$q^{lA}$ together with their superpartners becomes infinitely heavy and decouple, $k=1,...,K$, $l=(K+1),...,N$ and $A=1,...,K$,
$B=(K+1),...,N$. As a result in this limit our SQCD with gauge group U$(N)$ and $N_f=2N$ flavors splits in two non-interacting SQCDs with gauge groups U$(K)$ and $N_f=2K$ flavors. Eventually we will put $K=2$ so the starting point of our interpolation process will be \ntwo SQCD with the gauge group U(2) and $N_f=4$ quarks, which as we mentioned in the Introduction supports the critical non-Abelian vortex string.

 The final point of our interpolation process is the limit $M=0$. In this limit our theory
for $K=2$ becomes massless \ntwo SQCD with gauge group U(4) and $N_f=8$ quark flavors.

\subsection{World-sheet sigma model}
\label{sec:wcp}

The presence of the color-flavor locked group SU$(N)_{C+F}$ in 4D \ntwo SQCD with gauge group U$(N)$ is the reason for the formation of 
non-Abelian vortex strings \cite{HT1,ABEKY,SYmon,HT2}.
The most important feature of these vortices is the presence of the  orientational  zero modes.
As was already mentioned, in \ntwo SQCD these strings are 1/2 BPS saturated and preserve \ntwot supersymmetry on the world sheet.
Their tension is determined exactly by FI parameter,
\beq
T=2\pi \xi
\label{ten}
\eeq 

Let us briefly review the model emerging on the world sheet
of the non-Abelian  string \cite{SYrev}.

The translational moduli fields  are described by the Nambu-Goto action and  decouple from all other moduli. Below we focus on
 internal moduli.

If $N_f=N$  the dynamics of the orientational zero modes of the non-Abelian vortex, which become 
orientational moduli fields 
 on the world sheet, are described by 2D
\ntwot supersymmetric ${\mathbb{CP}}(N-1)$ model.

If one adds additional quark flavors, non-Abelian vortices become semilocal --
they acquire size moduli \cite{AchVas}.  
In particular, for the non-Abelian semilocal vortex in U($N$) \ntwo SQCD with $2N$ flavors,  in 
addition to  the complex orientational moduli  $n^i$ (here $i=1,...,N$), we must add $N$ complex size moduli   
$\rho^j$ (where $j=(N+1),...,2N$), see \cite{HT2,HT1,AchVas,SYsem,Jsem,SVY}. 

The effective theory on the string world sheet is a two-dimensional \ntwot supersymmetric \wcpN model, see review 
\cite{SYrev} for details. This model 
can be  defined  as a low energy limit of the  U(1) gauge theory \cite{W93}. The fields $n^{i}$ and $\rho^j$ have
charges  $+1$ and $-1$ respectively with respect to the  U(1) gauge field.
The bosonic part of this gauge linear  sigma model (GLSM) action reads
\begin{equation}
\begin{aligned}
	&S = \int d^2 x \left\{
	\left|\nabla_{\alpha} n^{i}\right|^2 
	+\left|\widetilde{\nabla}_{\alpha} \rho^j\right|^2
	-\frac1{4e_0^2}F^2_{\alpha\beta} + \frac1{e_0^2}\,
	\left|\pt_{\alpha}\sigma\right|^2 + \frac1{2e_0^2}\,D^2
	\right.
	\\[3mm]
	&-
	\left|\sqrt{2}\sigma +m_i\right|^2 \left|n^{i}\right|^2 - \left|\sqrt{2}\sigma +m_j\right|^2 \left|\rho^j\right|^2
	+D\left(\left|n^{i}\right|^2-\left|\rho^j\right|^2 - {\rm Re}\,\beta \right)
	\\[3mm]
	& \left.
	  - \frac{\vartheta}{2\pi}F_{01}
	\right\}, \quad \alpha,\beta=1,...,2\,,\quad i=1,...,N, \quad j=(N+1),...,2N,
	\end{aligned}
\label{wcpNN}
\end{equation}
where 
\begin{equation}
	\nabla_{\alpha}=\pt_{\alpha}-iA_{\alpha}\,,
	\qquad 
	\widetilde{\nabla}_{\alpha}=\pt_{\alpha}+iA_{\alpha}\,.	
	\label{cov_derivatives}
\end{equation}
The complex scalar $\sigma$ is a superpartner of the U(1) gauge field $A_{\alpha}$ and $D$ is the auxiliary field in the vector
supermultiplet.These fields can be written in terms of  the twisted chiral superfield $\Sigma$ \cite{W93}~\footnote{Here spinor indices are written as subscripts, say 
	$\theta^L=\theta_R$, $\theta^R= -\theta_L$. We also defined the twisted measure 
	$d^2 \tilde{\theta} = \frac12\,d \bar{\theta}_{L} d\theta_{R}$ to ensure that 
	$\int d^2 \tilde{\theta}\, \tilde{\theta}^2 = \int d \bar{\theta}_{L} d\theta_{R}\, \theta_R \bar{\theta}_L=1$.} 
\beq
\Sigma = \sigma +\sqrt{2}\theta_R \bar{\lambda}_L -\sqrt{2}\bar{\theta}_L \lambda_R
+\sqrt{2}\theta_R\bar{\theta}_L (D-iF_{01}).
\label{Sigma}
\eeq
The low energy limit in this model corresponds to $e_0^2\to \infty$ when components of the vector supermultiplet  classically
decouple due to the Higgs mechanism.

The complexified inverse coupling in \eqref{wcpNN}   
\begin{equation}
	\beta = {\rm Re}\,\beta + i \, \frac{\vartheta}{2 \pi} \,.
\label{beta_complexified}	
\end{equation}
is defined via  2D FI term  (twisted superpotential)
\beq
-\frac{\beta}{2}\,\int d^2 \tilde{\theta}\sqrt{2}\,\Sigma = - \frac{\beta}{2}\,(D-iF_{01})
\label{Sigma_sup}.
\eeq

Twisted masses $m_i$ and $m_j$ of fields $n^i$ and $\rho^j$ in \eqref{wcpNN} coincide with quark masses of 4D SQCD, namely with masses
$m_A$ of the first $N$ flavors, $A=1,...,N$ and ''extra'' flavors, $A=(N+1),...,2N$ respectively.

In the massless limit the number of real  bosonic degrees of freedom in the model \eqref{wcpNN} defines the dimension of its target space (Higgs branch), given by 
\beq
{\rm dim}_\mathbb{R}{\cal H} =4N-1-1=2\,(2N-1),
\label{dimH}
\eeq
where $4N$ is the number of  real degrees of freedom of $(n^i,\rho^j)$ fields  and we subtract one real $D$-term constraint 
\beq
|n^{i}|^2-|\rho^j|^2 = {\rm Re}\,\beta,
\label{D-term}
\eeq
in the limit $e_0^2\to \infty$,  and one gauge phase is eaten by the Higgs mechanism.

On the quantum level, the coupling $\beta$ does not run in this theory because the sum  of charges of $n$ and $\rho$ fields vanishes.  Hence, it is superconformal in the limit of  zero masses. 
Therefore, its target space is Ricci-flat and ( being K\"ahler due to \ntwot supersymmetry) represents  a (noncompact) Calabi-Yau manifold,  see 
\cite{NVafa,Bouchard} for  reviews on toric geometry.

The dimension of the Higgs branch \eqref{dimH} determines the central charge of the 2D conformal field theory (CFT) of the 
CY manifold
\beq
\hat{c}_{CY} \equiv \frac{c_{CY}}{3} = {\rm dim}_\mathbb{C}{\cal H} = 2N-1,
\label{cCY}
\eeq
just equal to its complex dimension. 
In $N=2$ case these ${\rm dim}_\mathbb{R}{\cal H} = 2(2N-1)=6$ internal degrees of freedom can be combined with four translational moduli of the non-Abelian vortex to form a 10D target space of a critical superstring \cite{SYcstring,KSYconifold}. 

Consider now  the classical vacuum structure of the  \wcpN model \eqref{wcpNN}.  At ${\rm Re}\,\beta >0$ we have $N$ vacua
\beq
\sqrt{2}\sigma=-m_{i_0}, \qquad |n^{i_0}|^2={\rm Re}\,\beta, \qquad i_0=1,...,N.
\label{wcpN_vac}
\eeq
Fields $n^i$, $i\neq i_0$ and fields $\rho^j$ have  masses $|m_i-m_{i_0}|$ and $|m_j-m_{i_0}|$ respectively. The number of vacua stay intact in the quantum theory because it is protected by Witten index which is equal to $N$.

Let us  discuss how our interpolation process is seen in the world sheet theory. Consider the choice of masses given by \eqref{extra_masses}, \eqref{masses}. In the limit $M\to\infty$ we have $K$  vacua with $\sigma=0$ where one of $n^{i_0}$ , $i_0=1,...,K$ develop VEV.
In these vacua fields $n^i$ with $i=(K+1),...,N$ and $\rho^j$, $j=(N+K+1),...,2N$ become infinitely heavy (with mass $|M|$) and decouple. The model \eqref{wcpNN} reduces to the  \wcpK model with massless fields  $n^i$, $i=1,...,K$ and $\rho^j$, 
$j=(N+1),...,(N+K)$. For $K=2$ the central charge $\hat{c}_{CY} =3$ and the \wcpt model becomes a sigma model on the conifold $Y_6$. In this case the non-Abelian vortex becomes a critical superstring.
This will be the starting point of our interpolating process. 

The final point corresponds to the limit $M=0$. The \wcpf model  is still conformal, but does not have the right central charge. We will show that the quantum world sheet model for 
$N=4$ case is not given by \wcpf model. As we already mentioned we will find the true string vacuum solving effective gravity equations.

Note, that in the limit $M\to\infty$ we could also consider another $K$  vacua with $\sqrt{2}\sigma=-M$. This would give 
another \wcpK model as a starting point. In 4D this is associated with U$(N)$ SQCD reducing to two non-interacting SQCDs with
 U$(K)$
gauge groups, see the previous subsection. Below for definiteness we will consider  the first option above.

To conclude this subsection we note that at $\beta=0$ the world sheet \wcpN model develop the Coulomb branch with arbitrary value of $\sigma$
in the massless limit. This can be shown  using the exact twisted superpotential for the \wcpN models as a function of twisted superfield  $\Sigma$.
This exact twisted superpotential is a generalization \cite{HaHo,DoHoTo}
of the CP($N-1$) model superpotential \cite{W93,AdDVecSal,ChVa,Dorey} of the Veneziano-Yankielowicz  type \cite{VYan}. 
The vacuum equation for $\sigma$ obtained by differentiating of this superpotential with respect to $\sigma$ reads \cite{GIMMY}
 \begin{equation}
	\prod_{i=1}^{N}\left(\sqrt{2} \, \sigma + m_i \right) 
		= e^{- 2 \pi \beta} \, \prod_{j = N+1}^{2N} \left(\sqrt{2} \, \sigma + m_j \right). \,
\label{2D_equation}	
\end{equation}
At generic values of masses it gives just $N$ distinct vacua with certain fixed values of $\sigma$. In the limit 
$m_i=m_j=0$ one gets
\beq
\sigma^N = e^{- 2 \pi \beta} \,\sigma^N.
\label{sigmaeq}
\eeq
with the $N$-degenerate vacuum solution $\sigma=0$ for any nonvanishing $\beta$. This means that fields $n$ and $\rho$
remain massless (see two first terms in the second
line in \eqref{wcpNN}) and live on the Higgs branch of the theory. However, for both $\beta=0$ and vanishing twisted masses the complex scalar $\sigma$ can have arbitrary value making $n^i$ and $\rho^j$ massive. This solution describes the Coulomb branch, which opens up at $\beta$ = 0. As was shown in \cite{GIMMY} this Coulomb branch can be effectively described in terms of \ntwo Liouville theory. 

Note also that the value $\beta=0$ at strong coupling is exactly what we are interested in for $N=2$ case.  The so-called ''thin string conjecture'' put forward in \cite{SYcstring,KSYconifold} implies that only at  $\beta=0$  we expect that the solitonic string-gauge duality works and the world sheet \wcpt model defines the right string theory for the critical non-Abelian vortex in \ntwo 4D SQCD.

\subsection {Massless 4D baryon}
\label{conifold}

In this section we consider the conifold case taking $N=2$ in the massless \wcpN model \eqref{wcpNN} and briefly review the only 4D massless state found in the string theory of the critical non-Abelian vortex \cite{KSYconifold}. It is associated 
with the deformation of the conifold complex structure. 
 As was already mentioned, all other massless string modes  have non-normalizable wave functions over the conifold. In particular, the 4D graviton associated with a constant wave
function over the conifold $Y_6$ is
absent as expected \cite{KSYconifold}. 

We can construct the U(1) gauge-invariant ``mesonic'' variables
\beq
w^{ij}= n^i \rho^j, \qquad i=1,2, \qquad j=3,4.
\label{w}
\eeq
These variables are subject to the constraint
\beq
{\rm det}\, w^{ij} =0. 
\label{coni}
\eeq

Equation (\ref{coni}) defines the conifold $Y_6$.  
It has the  Ricci-flat K\"ahler metric and represents a noncompact
 Calabi-Yau manifold \cite{Candel,NVafa,W93}. It is a cone which can be parametrized 
by the noncompact radial coordinate 
\beq
\widetilde{r}^{\, 2} = {\rm Tr}\, \bar{w}w\,
\label{tilder}
\eeq
and five angles, see \cite{Candel}. Its section at fixed $\widetilde{r}$ is $S_2\times S_3$.

At $\beta =0$ the conifold develops a conical singularity, so both spheres $S_2$ and $S_3$  
can shrink to zero.
The conifold singularity can be smoothed out
in two distinct ways: by deforming the K\"ahler form or by  deforming the 
complex structure. The first option is called the resolved conifold and amounts to keeping
a nonzero value of $\beta$ in (\ref{D-term}). This resolution preserves 
the K\"ahler structure and Ricci-flatness of the metric. 
If we put $\rho^K=0$ in \eqref{D-term} we get the $\mathbb{CP}(1)$ model with the sphere $S_2$ as a target space
(with the radius $\sqrt{\beta}$).  
The resolved conifold has no normalizable zero modes. 
In particular, 
the K\"ahler modulus $\beta$  which becomes a scalar field in four dimensions
 has a non-normalizable (quadratically divergent) wave function over the 
$Y_6$ and therefore is not dynamical \cite{KSYconifold}.  

If $\beta=0$ (i.e. exactly when the the Coulomb branch opens up) another option exists, namely a deformation 
of the complex structure \cite{NVafa}. 
It   preserves the
K\"ahler  structure and Ricci-flatness  of the conifold and is 
usually referred to as the deformed conifold. 
It  is defined by the deformation of Eq.~(\ref{coni}), namely,   
\beq
 {\rm det}\, w^{ij} = b\,,
\label{deformedconi}
\eeq
where $b$ is a complex parameter.
Now  the sphere $S_3$ can not shrink to zero, its minimal size is determined by $b$. 

We see that the resolved conifold corresponds to the Higgs branch of the GLSM \eqref{wcpNN} at
$N = 2$, while the deformed conifold is associated with the Coulomb branch
of this theory, which opens up at $\beta = 0$ \cite{GIMMY}.

The modulus $b$ becomes a 4D complex scalar field. The  effective action for  this field was calculated in \cite{KSYconifold}
using the explicit metric on the deformed conifold  \cite{Candel,Ohta,KlebStrass},
\beq
S_{{\rm kin}}(b) = T\int d^4x |\pt_{\mu} b|^2 \,
\log{\frac{\widetilde{R}_{\rm IR}^2}{|b|}}\,,
\label{Sb}
\eeq
where $\widetilde{R}_{\rm IR}$ is the  maximal value of the radial coordinate $\widetilde{r}$  introduced as an infrared (IR)regularization of the 
logarithmically divergent $b$-field  norm. Here the logarithmic integral at small $\widetilde{r}$ is cut off by the minimal size of $S_3$, which is equal to $|b|$.

We see that the norm of
the  modulus $b$ turns out to be  logarithmically divergent in the infrared \cite{Strom,KSYconifold}.
Such states at the borderline between normalizable 
and non-normalizable modes are considered as physical states ``localized'' in the 4D. 

 The field $b$  being massless can develop a VEV. Thus, 
we have a new Higgs branch in 4D \ntwo SQCD which is developed only for the critical value of 
the 4D coupling constant $\tau_{SW}=1$ associated with $\beta=0$ \cite{IYcorrelators}. 

In \cite{KSYconifold} the massless state $b$ was interpreted as a baryon of 4D \ntwo QCD. Its charge with respect to the baryonic U(1)$_B$ symmetry in \eqref{c+f} is $Q_B(b)=2$ \cite{KSYconifold}.

To conclude this section, we make a comment on non-normalizable strings modes.
As was mentioned above  most of  string modes have non-normalizable wave
functions over the conifold, i.e. they are not localized in 4D and cannot
be interpreted as dynamical states in 4D theory. Technically this happens because infinite normalization factor over the internal space appears in 4D kinetic terms for these states making them non-dynamical. These modes play a role of coupling constants in 4D theory, see \cite{GukVafaWitt} where the nature of non-normalizable string modes which appear upon Calabi-Yau compactifications was discussed. The example of such a coupling constant in the theory at hand is the inverse coupling $\beta$ of the world sheet sigma model \eqref{wcpNN} (it is related to 4D gauge coupling \cite{IYcorrelators}).  As we discussed above it corresponds to the K\"ahler form modulus of the conifold and has quadratically non-normalizable wave function.

\section {\ntwo Liouville theory from \wcpN model}
\label{wcp=liouville}
\setcounter{equation}{0}

In this section we first briefly review the derivation of the \ntwo Liouville theory from the world sheet \wcpN model at $\beta=0$
\cite{GIMMY} and then consider its deformation upon switching on masses in the \wcpN model.

\subsection{Massless theory}

Consider, first, the massless \wcpN model \eqref{wcpNN} in the large $N$  limit, $N\to\infty$. As we discussed in Sect.~\ref{sec:wcp} at $\beta=0$ the complex scalar 
$\sigma$ can take arbitrary values on the Coulomb branch of the theory.
For $\sigma\neq 0$ this makes the fields $n$ and $\rho$ massive, and one can integrate them out.  For both non-supersymmetric
and \ntwot supersymmetric \cpn models this was done by Witten \cite{W79} (see also \cite{SYhetN}).
He showed that the bare gauge coupling $e_0^2$
 taken to be infinite in the classical limit is renormalized at one loop
and becomes finite. This means that the U(1) gauge field introduced as an
auxiliary field in the GLSM formulation acquires a finite kinetic term and
becomes physical.

Almost the same calculation for \wcpN model gives the effective action for the vector multiplet, see \cite{GIMMY}. Focusing here on the most important kinetic term for $\sigma$ we get
\beq
S_{\rm eff}^{\sigma} = \int d^2 x  \frac1{e^2}|\pt_{\alpha}\sigma|^2, 
\label{eff_action_sigma}
\eeq
where the classical gauge coupling $e^2_0$ is corrected by the one loop contribution
\beq
\frac1{e^2} = \left.\left( \frac1{e_0^2} + \frac{2N}{4\pi}\,\frac{1}{2|\sigma|^2}\right)\right|_{e^2_0\to\infty} = \frac{2N}{4\pi}\,\frac{1}{2|\sigma|^2}\,.
\label{e}
\eeq
The wave function renormalization  comes from $n^i$ and $\rho^j$ fields (with their fermionic superpartners )  propagating  in the loop. 
The loop integral is finite in the ultraviolet (UV) region and is saturated in the IR region at momenta of order of $n$ and $\rho$ ``mass''
$\sqrt{2}|\sigma|$, see \eqref{wcpNN}~\footnote{We put here ``mass'' in quotation marks, since in 2D theory $\sigma$ does not have a definite VEV, instead the ground state wave function is spread over the whole Coulomb branch, cf. \cite{W44}. }.
The loop graph  contains two vertices, each proportional to the electric charge of a given $n$ or $\rho$ field (equal to $\pm1$). Therefore, this graph is proportional to the sum of
squires of these electric charges, which is equal to $2N$.
The result \eqref{e} gives the leading term in the $1/N$ expansion. We have
\beq
S_{\rm eff}^{\sigma}=\frac{2N}{4\pi}\int d^2 x \; \frac1{2}\,\frac{|\pt_{\alpha}\sigma|^2}{|\sigma|^2} 
\label{tube}
\eeq
with the tube metric~\footnote{The metric looks singular, but actually 
there is no singularity at the origin  \cite{W44}.}. 
Making a change of variables
\beq
\sigma= e^{-\frac{\phi + iY}{Q}},
\label{sigma}
\eeq
where we parametrized the modulus of $\sigma$ by  the real scalar field $\phi$ (which will be the Liouville field) and its phase by the real compact scalar $Y$ with the periodicity condition
\beq
Y+2\pi Q \sim Y
\eeq
we arrive to the bosonic part of the effective action 
\beq
S_{\rm eff}=\frac{1}{4\pi}\int d^2 x \;\left( \frac1{2}\,(\pt_{\alpha}\phi)^2 + \frac1{2}\,(\pt_{\alpha}Y)^2\right)  
\label{free_action}
\eeq
where the radius of the compact dimension 
\beq
Q\stackreb{N\to\infty}{\approx}\sqrt{2N}.
\label{Q_largeN}
\eeq
Other components of the vector supermultiplet can be considered similarly, see \cite{GIMMY}. For example, the U(1) gauge field has no physical degrees of freedom in two dimensions and can be integrated out together with the $D$-field, see \cite{GIMMY} for details.

Repeating the above calculation on the curved world sheet we can restore the background charge  of the Liouville field 
\cite{GIMMY}
\beq
S_{\rm eff}=\frac{1}{4\pi}\int d^2 x\sqrt{h}
 \;\left(\frac1{2}\,h^{\alpha\beta}(\pt_{\alpha}\phi\pt_{\beta}\phi  +\pt_{\alpha}Y\pt_{\beta}Y)
 -\frac{Q}{2}\phi\, R^{(2)} \right),
\label{Liouville}
\eeq
where $h_{\alpha\beta}$ is the world sheet metric, $R^{(2)}$ is the world sheet Ricci scalar and  $h={\rm det} (h_{\alpha\beta})$. 

This  is exactly the bosonic part of the  \ntwo Liouville action, see \cite{Nakayama} for a review. Note the linear  dilaton in 
\eqref{Liouville},
\beq
\Phi =-\frac{Q}2\,\phi
\label{linear_dilaton}
\eeq
with the  background charge $Q$ for the  Liouville field $\phi$ 
 which coincides with the radius of the compact dimension (as it should in the \ntwo Liouville theory). In the large $N$ approximation $Q$ is given by \eqref{Q_largeN}.

The action in \eqref{Liouville} leads to the following holomorphic stress tensor of the bosonic part of the theory 
\beq
T= -\frac12\,\left[(\pt_z \phi)^2 + Q\, \pt_z^2 \phi + (\pt_z Y)^2\right]. 
\label{T--}
\eeq

The \ntwo Liouville interaction superpotential (see \cite{Nakayama}) comes from the 2D FI term \eqref{Sigma_sup} in the \wcpN model,
\beq
L_{int}= \mu\,\int d^2 \tilde{\theta}\,\Sigma = \mu\,\int d^2 \tilde{\theta}\,e^{-\frac{\phi +iY}{Q}}
\label{Liouville_sup},
\eeq
where we use parametrization \eqref{sigma} and promote scalars $\phi$ and $Y$ to (twisted) chiral superfields, see \cite{GIMMY}
for details~\footnote{The fact that the Liouville interaction is given by a twisted superpotential  is just a matter of conventions since there are no untwisted chiral fields in the effective theory.}.

This superpotential is a marginal deformation of the \ntwo Liouville theory \eqref{Liouville}. The  conformal dimension of
$\sigma$ is 
\beq
\Delta (\sigma=e^{-\frac{\phi +iY}{Q}})= \left(\frac12,\,\frac12\right),
\label{dim_sigma}
\eeq
which can be easily checked using the stress tensor \eqref{T--}.

The above outlined equivalence of the Coulomb branch of \wcpN model and \ntwo Liouville theory obtained in the large $N$ approximation can be promoted to the exact equivalence. It was argued in \cite{GIMMY} that the $\sigma$-dependence of the effective action \eqref{tube} is fixed on dimensional grounds and integrating fields $n^i$ and $\rho^j$ exactly rather then in the large $N$ approximation we would arrive to the same action \eqref{tube} with the coefficient $2N$ replaced by the exact coefficient $Q^2(N)$. To find the exact dependence of $Q^2(N)$ on $N$ we can demand that central charges of both CFTs (\wcpN model and \ntwo Liouville theory) should coincide. The central charge of the \ntwo Liouville theory is
\beq
\hat{c}_L= 1+Q^2.
\label{c_L}
\eeq
Requiring that it should be equal to the central charge $\hat{c}_{CY}$ \eqref{cCY} gives the exact relation
\beq
Q(N)=\sqrt{2(N-1)},
\label{Q}
\eeq
which reduces to \eqref{Q_largeN} in the large $N$ approximation.

 Note also that as we already mentioned for the  case $N=2$ the Coulomb branch of \wcpt model is associated with the deformed conifold. Therefore, the coefficient $\mu$ in front of the marginal deformation \eqref{Liouville_sup}
should be identified with the conifold complex structure parameter $b$ \cite{GivKut,GivKutP,GIMMY}, \footnote{The unit
 power of $b$ in the r.h.s. of \eqref{mu=b} was fixed in \cite{SYlittles} using the baryonic U(1) symmetry.}
\beq
\mu\sim b.
\label{mu=b}
\eeq
 On the CY side 
parameter $b$ smoothies  the conifold singularity at small $\widetilde{r}$, i.e. provides a UV regularization. In the Liouville theory the Liouville superpotential (the Liouville wall) at nonzero $\mu$ also provides a UV  regularization preventing
field $\phi$ from penetrating to the region of large negative values.   With the identification \eqref{mu=b}   the conifold complex structure modulus which was not seen in the GLSM description \eqref{wcpNN} becomes manifest in the Liouville description. 

To conclude this subsection, we note that  the dilaton has a linear dependence on the Liouville coordinate $\phi$, see \eqref{linear_dilaton}. Therefore, the string coupling constant $g_s=e^{\Phi}$ would become large at large negative $\phi$. On the other hand  at nonzero $b$ the Liouville wall 
prevents field $\phi$ from penetrating to the region of large negative values. In fact, the maximum value of the string coupling is 
$g_s\sim 1/|b|$ for $Q=\sqrt{2}$. In this paper we keep $b$
large to ensure that the string coupling is small and the string
perturbation theory is reliable, see \cite{GivKut,SYlittmult}. In particular, we can use
the tree-level approximation to obtain the string spectrum.

In terms of 4D SQCD taking $ b$ large
means moving along the Higgs branch far away from the origin.

\subsection{Primary operators }
\label{sec:vertices}

In this subsection we review primary operators in the \ntwo Liouville theory. For $N=2$ case they describe physical string states interpreted as  hadrons in 4D SQCD, see \cite{SYlittles} for details.

Primary operators for  the  \ntwo Liouville theory are constructed in \cite{GivKut}, see also 
\cite{GivKutP,MukVafa}. For large 
positive $\phi$  (where the Liouville interaction is small) primaries take the form
\beq
T_{j;m_L,m_R} = e^{Q\left[j\phi + i(m_L Y_L - m_R Y_R)\right]},
\label{vertex}
\eeq
where we split  $Y$ into left and right-moving parts. Parameters $m_L$ and $m_R$ for 
  left-moving and right-moving sectors are given by
\beq
m_L= \frac12(n_1+kn_2), \qquad  m_R= \frac12(n_1-kn_2),
\label{m}
\eeq
where $n_2$ and $n_1$ are integers corresponding to momentum and winding numbers along the compact dimension $Y$.

The primary operator \eqref{vertex} is related  to the wave 
function on the target space   as follows:
\beq
T_{j;m_L,m_R} = g_s \Psi_{j;m_L,m_R}(\phi,Y)\,,
\label{Psi}
\eeq
where the string coupling $g_s= e^{\Phi}$  depends on $\phi$, see \eqref{linear_dilaton}. Thus, 
\beq
\Psi_{j;m_L,m_R}(\phi,Y) \sim e^{\sqrt{2}(j+\frac{1}{2})\phi + i\sqrt{2}(m_L Y_L +m_R Y_R)}\,.
\eeq
We will look for string states with normalizable along the noncompact Liouville dimension wave functions. These states are localized in 4D and can be interpreted as hadrons in 4D SQCD. The 
condition for the  states to have  normalizable wave functions reduces to 
\beq
j\le -\frac12\,.
\label{normalizable}
\eeq
We include  the case $j=-\frac12$
which is at the borderline between normalizable and non-normalizable states. 

The conformal dimension of the primary operator \eqref{vertex} is 
\beq
\Delta_{j,m} = \frac{Q^2}{2}\left\{m^2 - j(j+1)\right\}  .
\label{dimV}
\eeq

Unitarity implies that the conformal dimension \eqref{dimV} should be positive,
\beq
\Delta_{j,m}> 0\,.
\label{Deltapositive}
\eeq
Moreover, to ensure that conformal dimensions of left and right-moving parts of the vertex operator
\eqref{vertex} are the same we impose that $m_R=\pm m_L$.

The \ntwo Liouville theory has  a mirror description  \cite{HoriKapustin}  in terms of a supersymmetric version of the two-dimensional
black hole with the cigar geometry \cite{Wbh}, which is the  \ntwo SL($2, \mathbb{R}$)/U(1) coset WZNW theory 
\cite{GVafa,GivKut,MukVafa,OoguriVafa95} 
at the level
\begin{equation}
	k =\frac{2}{Q^2}\,.
\label{k_Q_relation}
\end{equation}
 of the Ka\v{c}-Moody algebra. 

The spectrum of the allowed values of $j$ and $m$ in \eqref{vertex} was  exactly determined  using the Ka\v{c}-Moody algebra
for the mirror description of the theory  in \cite{MukVafa,DixonPeskinLy,Petrop,Hwang,EGPerry}, 
see \cite{EGPerry-rev} for a review. Both discrete and continuous representations were found. Parameters $j$
and $m$ determine the global quadratic Casimir operator and the projection of the spin  on the third axis,
\beq
J^2\, |j,m\rangle\, = -j(j+1)\,|j,m\rangle, \qquad J^3\,|j,m\rangle\, =m \,|j,m\rangle .
\eeq

We have 

(i) {\em Discrete representations} with
\beq
j=-\frac12, -1, -\frac32,..., \qquad m=\pm\{j, j-1,j-2,...\}.
\label{discrete}
\eeq

(ii) {\em Principal} continuous representations with
\beq
j=-\frac12 +is, \qquad m= {\rm integer} \quad {\rm or} \quad m= \mbox{ half-integer},
\label{principal}
\eeq
where $s$ is a real parameter.

We see that discrete representations include  normalizable and borderline-normalizable states localized near the tip of the cigar.
  This nicely matches our qualitative expectations.
	
Consider now the \ntwo Liouville theory with $N=2$, $Q =\sqrt{2(N-1)}= \sqrt{2}$.  This corresponds to $k=1$ in the mirror description on the cigar. Take the primary operator \eqref{vertex} with $j=-1/2$ and $m_L=\pm 1/2$ . Its conformal dimension 
is 
\beq
\Delta_{j=-\frac12,m=\pm \frac12 } =\frac12 
\label{Delta_b}
\eeq
(see \eqref{dimV}), so it is marginal and describes a massless string state in 4D. As was noticed in \cite{SYlittles} this massless state corresponds to the complex structure modulus $b$ for the string compactification on the conifold. Two possible values of $m=\pm 1/2$ corresponds to two real degrees of freedom of the complex scalar field $b$.
	The associated string state has a logarithmically normalizable wave function over the conifold in terms of the radial coordinate $\widetilde{r}$ \cite{Strom,KSYconifold}, see\eqref{Sb}. On the Liouville side this corresponds to the borderline normalization of the massless state \eqref{vertex} with $j=-\frac12$, $m=\pm \frac12$, see \cite{SYlittles} for details.
	
The discrete spectrum \eqref{discrete} gives rise to physical hadron states in 4D SQCD. In particular, the mass spectrum of massive 4D states created by vertex operators \eqref{vertex} with $j=-\frac12$ has the form \cite{SYlittles},
\beq
\frac{M_{\rm 4D}^2}{8\pi T}= \Delta_{j,m} -\frac12,
\label{4D_masses}
\eeq
where $Q=\sqrt{2}$.

To conclude this subsection, let us make a comment on the   principal continuous representation
\eqref{principal}  of string states, which represents plane waves in the non-compact Liouville dimension. In \cite{SYlittles,IYcorrelators} it was  suggested an interpretation of these states: they correspond to multiparticle states associated with decay of normalizable 4D states. This interpretation is motivated by
the observation that spectra of continuous states for half-integer $m$ start from thresholds given by
masses of (borderline) normalizable states. This issue however needs future clarification.

\subsection{Mass deformation}
\label{sec:mass_deform}

Now consider \wcpN model \eqref{wcpNN} with nonzero twisted masses   starting with  the large $N$ approximation. Integrating out $n^i$ and $\rho^j$ fields at $\beta =0$
we instead of \eqref{tube} get 
\beq
S_{\rm eff}^{\sigma}=\frac{1}{4\pi}\int d^2 x \, \sum^{2N}_{A=1}\frac{|\pt_{\alpha}\sigma|^2}{|\sqrt{2}\sigma + m_A|^2} 
= \frac{1}{4\pi}\int d^2 x \, \frac12\,\frac{|\pt_{\alpha}\sigma|^2}{|\sigma |^2}\sum^{2N}_{A=1}\frac{1}{\left| 1+ 
\frac{m_A}{\sqrt{2}\sigma}\right|^2
\label{tube_deformed}}
\eeq
for  the effective action of the field $\sigma$.

Consider the choice of masses given by \eqref{extra_masses}, \eqref{masses}.  $K$ first $n$ fields are massless,  $K$ last 
$n$ fields are massive with the same mass $M$, $N=2K$, while masses of $\rho$ fields are equal to masses of $n$ fields, see \eqref{extra_masses}. The action \eqref{tube_deformed} takes the form 
\beq
S_{\rm eff}=\frac{1}{4\pi}\int d^2 x \; g_{cl}(\phi,Y)\,\left( \frac1{2}\,(\pt_{\alpha}\phi)^2 + \frac1{2}\,(\pt_{\alpha}Y)^2\right)  
\label{deformed_action}
\eeq
where we use the parametrization \eqref{sigma}, and  the ''classical'' warp factor of the target space metric is
\footnote{ A comment on dimensions is in order. The canonical dimension of $\sigma$ is unity. We can introduce dimensionless field $\sigma' =\sigma/\sqrt{4\pi T}$ and use parametrization \eqref{sigma} for $\sigma'$. Then dimensionless $M'=M/\sqrt{4\pi T}$
appears in \eqref{cl_warp}. Below we consider dimensionless quantities dropping primes to simplify  notations. We also remind that $2\pi T=1/\alpha'$.}
\beq
g_{cl}(\phi,Y)= 1+ \frac{1}{\left|1+\frac{M}{\sqrt{2}}\,e^{\frac{\phi+iY}{Q}}\right|^2},
\label{cl_warp}
\eeq
while the radius of the compact dimension 
\beq
Q\stackreb{K\to\infty}{\approx}\sqrt{2K}.
\label{Q_largeK}
\eeq
We also drop the dilaton term in \eqref{deformed_action} which we will restore later.

As we already mentioned non-zero twisted masses break conformal invariance in the world sheet model model \wcpN. Therefore,
we cannot use the mass-deformed model \eqref{deformed_action} for the string quantization. To find the true string vacuum  we will solve the effective supergravity equations of motion in the next section. To find this solution we will use the expansion of the classical target space metric in \eqref{deformed_action} just as initial conditions at  $\phi\to \infty$ where the deformation is small. Namely, expanding \eqref{deformed_action} at large $\phi$ we write the classical warp factor \eqref{cl_warp}
as
\beq
g_{cl}(\phi,Y)\approx 1+ \frac{2}{|M|^2}\,e^{-\frac{2\phi}{Q}} +\cdots,
\label{warp_bc}
\eeq
and use this expression in the next section as  initial conditions for the true quantum warp factor at $\phi\to\infty$. 
It just shows the initial ''position'' and the ''velocity'', in which the true quantum metric is pushed by the mass
deformation, while the true ''trajectory'' should be found by solving the gravity equations of motion.

As initial conditions for the dilaton at large $\phi$ we use the linear dilaton in \eqref{linear_dilaton},
\beq
\Phi \approx -\frac{Q}2\,\phi +\cdots.
\label{dilaton_bc}
\eeq

The starting point of our interpolation procedure is the limit $M\to\infty$ where the world sheet theory reduces to \wcpK model and its Coulomb branch at $\beta=0$ is given by the \ntwo Liouville theory. As we mentioned above this equivalence is exact in $K$ and we relax the large $K$ condition using exact expression 
\beq
Q(K)=\sqrt{2(K-1)},
\label{Q(K)}
\eeq 
see \eqref{Q}. For the case $K=2$ when non-Abelian vortex string become a critical superstring this gives 
\beq
Q=\sqrt{2}.
\label{Q_coni}
\eeq
We also assume that boundary conditions \eqref{warp_bc} and \eqref{dilaton_bc} depend on $K$ via $Q(K)$ \eqref{Q(K)} and extrapolate 
eqs. \eqref{warp_bc} and \eqref{dilaton_bc} to $K=2$.

\section {Solutions of gravity equations}
\label{sec:gravity}
\setcounter{equation}{0}

In this section we study effective gravity equations for the mass-deformed superstring background and find their solutions.

\subsection{The setup}

The bosonic part of the action of the type-II supergravity
in the string frame is given by
\beq
S= \frac1{2\kappa^2}\int d^D x \, \sqrt{-G}\,e^{-2\Phi}\,\left\{ R + 4G^{MN}\pt_M\Phi\pt_N\Phi_N +\cdots \right\},
\label{gravity_action}
\eeq
where $G_{MN}$ is the $D$-dimensional metric and we keep in \eqref{gravity_action} only the metric and the dilaton terms, $M,N=1,...,D$.
Here $2\kappa^2= (2\pi)^{(\frac{D}{2} -2)}g^2_s/T^{\frac{D-2}{2}}$.

Einstein's equations of motion following from the action \eqref{gravity_action} have the form
\beq
R_{MN}+ 2D_M D_N \Phi =0,
\label{Einstein}
\eeq
while the equation for the dilaton reads
\beq
R=4G^{MN}\pt_M\Phi\pt_N\Phi -4G^{MN}D_M D_N \Phi +p,
\label{dilaton_eq}
\eeq
where $p=\frac{D-10}{2}$ (in dimensionless units) is included if $D\neq 10$.	

We assume that our space-time is a direct product of the flat 4D Minkowski space and an internal space which has the non-trivial metric of the target space of the \ntwo deformed Liouville theory. Thus, $D=6$ and the ansatz for the internal metric is
\beq
ds^2_{{\rm int}} = g(\phi,Y) \left\{ d^2\phi + d^2 Y\right\}.
\label{int_metric}
\eeq
It is inspired by the calculation in Sec. \ref{sec:mass_deform}.

Let us  note that in the limit $M\to \infty$ equations of motion are satisfied by the flat internal metric with $g=1$ and the linear dilaton \eqref{linear_dilaton}. The Einstein's equation is satisfied because Ricci tensor is zero for the flat metric and covariant derivatives in \eqref{Einstein} reduce to  ordinary ones, so the second term in the l.h.s. of \eqref{Einstein}  gives zero on the linear dilaton. Eq. \eqref{dilaton_eq} is satisfied for for $Q=\sqrt{2}$ and $p=-2$, see \eqref{Q_coni}.

\subsection{Solutions to the gravity equations}
\label{sec:grav_sol}

In this section we find a solution to the gravity equations \eqref{Einstein} and \eqref{dilaton_eq} which satisfy initial conditions \eqref{warp_bc} and \eqref{dilaton_bc}. Flat Minkowski part of equations is trivial and decouples so we are left with gravity equations for the internal part.  Due to \ntwot supersymmetry the metric of the internal space is K\"ahler.
Therefore we introduce complex coordinates
\beq
s= \phi +iY, \qquad \bar{s}= \phi -iY.
\label{s}
\eeq
In terms of these coordinates the metric \eqref{int_metric} takes the form
\beq
ds^2_{{\rm int}} = g(s,\bar{s}) \,ds\,d\bar{s}, \qquad g_{\bar{s}s} =g_{s\bar{s}}= \frac12\,g(s,\bar{s}).
\label{int_metric_c}
\eeq

For this metric the only nonzero Christoffel symbols are
\beq
\Gamma_{ss}^{s}= \frac1{g}\pt_s g,\qquad \Gamma_{\bar{s}\bar{s}}^{\bar{s}}= \frac1{g}\pt_{\bar{s}} g
\label{Christoffel}
\eeq
and nonzero Ricci tensor components take the form
\beq
R_{s\bar{s}} = R_{\bar{s}s} =-\pt_s \pt_{\bar{s}} \ln{g},
\label{Ricci}
\eeq
while Ricci scalar reads
\beq
R=-\frac{4}{g}\,\pt_s \pt_{\bar{s}} \ln{g}.
\label{Ricci_scalar}
\eeq

Then Einstein equations \eqref{Einstein} with $s\bar{s}$ and $ss$ indices reduce to
\beq
-\pt_s \pt_{\bar{s}} \ln{g} +2\,\pt_s \pt_{\bar{s}} \Phi =0
\label{sbars_eq_0}
\eeq
and 
\beq
\pt_s\pt_s \Phi -  \frac1{g}\pt_s g\,\pt_s\Phi =0
\label{ss_eq_0}
\eeq
respectively, while the equation with $\bar{s}\bar{s}$ indices is just a complex conjugate of \eqref{ss_eq_0}.

The dilaton equation \eqref{dilaton_eq} reads
\beq
\frac{1}{g}\,\pt_s \pt_{\bar{s}} \ln{g} + \frac{4}{g}\,\pt_s  \Phi\pt_{\bar{s}} \Phi - \frac{4}{g}\,\pt_s \pt_{\bar{s}}\Phi 
+\frac{p}{4} =0.
\label{dilaton_eq_0}
\eeq

The solution to the equation \eqref{sbars_eq_0} has the form
\beq
\Phi = -\frac{Q}{4}\,(s+\bar{s}) + \frac12\, \ln{g},
\label{Phi}
\eeq
where we use initial conditions \eqref{dilaton_bc}. In principle, we can add to the r.h.s. of \eqref{Phi} arbitrary 
holomorphic function of $s$ plus its complex conjugate. However it is easy to see that this corresponds just to  a re-parametrization of of the variable $s$. We fix the gauge assuming that this function is zero.

Substituting \eqref{Phi} in the Einstein equation \eqref{ss_eq_0} and the dilaton equation \eqref{dilaton_eq_0} gives 
\beq
\pt_{s}^2 \ln{g} -(\pt_s \ln{g})^2 +\frac{Q}{2}\,\pt_s \ln{g}=0
\label{ss_eq}
\eeq 
and 
\beq
-\pt_{\bar{s}}\pt_s \ln{g} +\frac{Q^2}{4} +\frac{p}{4}\,g - \frac{Q}{2}\left(\pt_s \ln{g}+ \pt_{\bar{s}}\ln{g}\right)
+\pt_s \ln{g}\,\pt_{\bar{s}}\ln{g}=0
\label{dilaton_eq_g}
\eeq
respectively.

The equation \eqref{ss_eq} is a first order equation for the variable $\pt_s \ln{g}$ and admits separation of variables. One gets
\beq
\pt_s \ln{g}= \frac{Q}{2}\, \frac1{1-f(\bar{s})\,e^{\frac{Q}{2} s}},
\eeq
where $f(\bar{s})$ is a function of $\bar{s}$. To fix this function we use the complex conjugate of the equation \eqref{ss_eq}.
This gives
\beq
\pt_s \ln{g}= \pt_{\bar{s}} \ln{g}=\frac{Q}{2}\, \frac1{1- A\,e^{\frac{Q}{2} (s+\bar{s})}},
\label{pt_g}
\eeq
where $A$ is a constant. Integrating this equation we finally get
\beq
g= \frac1{1-\frac1{A}\,e^{-\frac{Q}{2} (s+\bar{s})}},
\label{g(s)}
\eeq
where we put the integration constant to zero using the initial condition $g=1$ at $s\to\infty$.
It is easy to check that this solution satisfies also the dilaton equation \eqref{dilaton_eq_g} for $Q=\sqrt{2}$ and $p=-2$.

The warp factor \eqref{g(s)} can be written  as
\beq
g(\phi)= \frac1{1-\frac1{A}\,e^{- Q \phi}}= \frac1{1-e^{- Q (\phi-\phi_0)}},
\label{g}
\eeq
while the solution for the dilaton takes the form
\beq
\Phi (\phi)= -\frac{Q}{2} \phi -\frac12\, \ln{\left(1-\frac1{A}\,e^{- Q \phi} \right)}=-\frac{Q}{2} \phi -\frac12\, 
\ln{\left[1-e^{- Q (\phi-\phi_0)}\right]},
\label{Phi_sol}
\eeq
where we use \eqref{Phi}. Here we introduce $\phi_0=-\frac1{Q}\ln{A}$.
 We see that the warp factor of the metric and the dilaton are functions of the Liouville field $\phi$ and do not depend on $Y$.

Observe now that precisely for our case $Q=\sqrt{2}$, which corresponds to  $K=2$ the warp factor \eqref{g} satisfy  initial conditions
\eqref{warp_bc} associated with the mass deformation. Now we can identify the parameter $A$ in terms of the mass $M$. We have 
\beq
A=\frac{M^2}{2}\, , \qquad \phi_0 = -\frac1{Q}\ln{\left(\frac{M^2}{2}\right)}.
\label{A_m}
\eeq

Note also that the first non-trivial term in the expansion of the warp factor \eqref{g} at large $\phi$ gives rise to the following
deformation operator
\beq
 (\pt_{z}\phi - i\pt_{z}Y) (\pt_{\bar{z}}\phi + i\pt_{\bar{z}}Y)  \,e^{- Q \phi}.
\label{non-chiral_deform}
\eeq
This operator has $j=-1$, $m=0$ and  is marginal with conformal dimension $\Delta =(1,1)$. It is  the bosonic part of  so called non-chiral marginal deformation of  \ntwo Lioville theory, see \cite{Nakayama} for a review. We see that \eqref{g}  and \eqref{Phi_sol} represent exact solution for the mass  deformation which infinitesimally is associated  with the non-chiral marginal operator \eqref{non-chiral_deform}.

Solutions \eqref{g}  and \eqref{Phi_sol} define the true quantum vacuum of the mass-deformed string theory. Namely, the  bosonic part of the mass-deformed \ntwo Liouville world sheet theory takes the form 
\beqn
 S_{\rm ws} &=&\frac{1}{4\pi}\int d^2 x \sqrt{h}\; \left\{g(\phi)\left[ \frac1{2}\,(\pt_{\alpha}\phi)^2 + \frac1{2}\,(\pt_{\alpha}Y)^2\right] 
\right.
\nonumber\\
 && \left.
+ \;\Phi(\phi)\,R^{(2)} + L_{int}\right\},
\label{deformed_Liouville}
\eeqn
where the metric warp factor $g(\phi)$ and the dilaton $\Phi(\phi)$ are given by \eqref{g}  and \eqref{Phi_sol}.
 Also we will show in the next section  that the Liouville superpotential \eqref{Liouville_sup} is still a marginal deformation of the theory therefore, $L_{int}$ in \eqref{deformed_Liouville} is not modified and is still given by \eqref{Liouville_sup}.

Thus, the action \eqref{deformed_Liouville} defines a continues family of CFTs with the same central charge $\hat{c}_{L}=3$ ( see 
\eqref{c_L}) parametrized by the mass parameter $M$ which 
we can use as  world sheet theories  for the string quantization.

\subsection{Scales of the deformed Liouville theory}

Let us discuss some properties of our solution.
The metric warp factor \eqref{g} develop a naked singularity at $\phi=\phi_0$,
\beq
g|_{\phi\to\phi_0} \approx \frac1{Q(\phi - \phi_0)}
\label{pole}
\eeq
with Ricci tensor  given by
\beq
R_{s\bar{s}} =- \frac{Q^2}{4}\, \frac{A\,e^{\frac{Q}{2} (s+\bar{s})}}{\left[1-A\,e^{\frac{Q}{2} (s+\bar{s})}\right]^2}\,
|_{\phi\to\phi_0} \approx \frac14\,\frac1{(\phi - \phi_0)^2}
\label{Ricci_sing}
\eeq
Thus, the geometry is defined only at $\phi>\phi_0$.

Note that our exact solution for the metric warp factor \eqref{g} has some qualitative similarity with the ''classical'' warp factor \eqref{cl_warp} obtained by integration out massive $n$ and $\rho$ fields in the \wcpN model.  Namely,  the classical warp factor \eqref{cl_warp} also has a singularity at $\phi=\phi_0$ if we take  $Y=\pi Q$, however the type of the singularity is different. Solving gravity equations of motion  allows us to find a true quantum  string vacuum which arises due to the mass deformation.

The  deformed \ntwo Liouville theory \eqref{deformed_Liouville} has two scales. The first one is associated with the Liouville wall (the superpotential \eqref{Liouville_sup}) which prevents field $\phi$ from penetrating to the region of large negative values.
The Liouville interaction becomes of order of unity at 
\beq
\phi_{\rm wall} \sim -\,Q\ln{\frac1{|b|}},
\label{phi_wall}
\eeq
where we used \eqref{mu=b} for $K=2$. The Liouville wall prevents $\phi$ from penetrating far below this value.

The second scale is associated with the singularity of the target space metric at $\phi=\phi_0$. 

As we start our interpolating process at $M\to\infty$, $\phi_0\to -\infty$ and is much smaller then $\phi_{\rm wall}$,  $\phi_0\ll \phi_{\rm wall}$ so the geometry is almost flat in the allowed region of $\phi$. The string spectrum associated with the Liouville world sheet theory found in \cite{SYlittles} describes hadrons of \ntwo SQCD with U(2) gauge group and $N_f=4$ quark flavors. As the  mass $M$ reduces the geometry gets deformed and at $\phi_0 \sim \phi_{\rm wall}$ ( $M^2\sim 1/|b|^2$)  we expect a transition  to the region of small $M$.

In the opposite limit $\phi_0\gg \phi_{\rm wall}$ in the region of small $M$  the effect of the Liouville wall can be neglected and the string background given by 
\eqref{g} and \eqref{Phi_sol}  determines  the string spectrum.  In this limit our string theory is expected to describe hadrons
of \ntwo SQCD with U(4) gauge group and $N_f=8$ quark flavors.

In the next section we take a first glance at the string spectrum  leaving its detail study for  a future work.

\section {A first glance at the string spectrum}
\label{sec:spectrum}
\setcounter{equation}{0}

In this section we develop an effective gravity approach which can be used  to study the string spectrum  associated with the mass-deformed \ntwo Liouville world sheet theory \eqref{deformed_Liouville}. In particular, we show that massless 4D baryon $b$ survive
the mass deformation.

\subsection{Tachyon equation}
\label{sec:tachyon_eq}

Primary tachyon vertex operators \eqref{vertex} can be described as scalar fields in the effective supergravity \eqref{gravity_action} \footnote{These states are, of course, not tachyonic in 4D, but we will use the standard terminology and refer to them as ''tachyons''.}. To take them into account we add the tachyonic term
\beq
S_{\rm tachyon}= \frac1{2\kappa^2}\int d^D x \, \sqrt{-G}\,e^{-2\Phi}\,\left\{-G^{MN}\pt_M \bar{T}_{j,m}\pt_N T_{j,m} 
+ |T_{j,m}|^2\right\}.
\label{tachyon_action}
\eeq
to the gravity action \eqref{gravity_action}, cf. \cite{Polch_90}.
This gives the equation for the tachyon field
\beq
D_N D^N T_{j,m} -2\pt_N\Phi\pt^N T_{j,m} + T_{j,m}=0
\label{T_eq_gen}
\eeq
and we neglect the back reaction of tachyons on the metric and the dilaton.

Dressing the tachyon with the dependence on the 4D coordinates 
\beq
e^{ip_{\mu}x^{\mu}}\, T_{j,m}
\label{T_dressed}
\eeq
we rewrite the tachyon equation \eqref{T_eq_gen} in the form
\beq
\frac4{g}\left\{\pt_s\pt_{\bar{s}} T_{j,m} -\pt_s \Phi \,\pt_{\bar{s}} T_{j,m} -\pt_{\bar{s}}\Phi \,\pt_s T_{j,m}\right\} 
+ \left(1+ \frac{M^2_{\rm 4D}}{4\pi T}\right)T_{j,m} =0,
\label{T_eq}
\eeq
where we used complex coordinates \eqref{s}. Here the mass squared of the physical state in 4D is 
\beq
M^2_{\rm 4D}= -p_{\mu}p^{\mu},
\eeq
where the Minkowski 4D metric with the diagonal entries $(-1,1,1,1)$ is used.

To conclude this subsection  let us solve the equation \eqref{T_eq}  in the limit $M\to\infty$ when $g=1$ and the dilaton is given by 
\eqref{linear_dilaton}. 
 In this limit the solution of the equation \eqref{T_eq} can be written in the form 
\beq
T_{j,m} = e^{Q[j\phi +imY]} = e^{Q[J_{+}s+J_{-}\bar{s}]},
\label{T_0}
\eeq
where 
\beq
J_{+}=\frac12(j+m), \qquad J_{-}=\frac12(j-m)
\label{J}
\eeq
and we consider, say,  the momentum modes along the compact dimension, $m\equiv m_L=-m_R$ for definiteness.
Substituting this to to the equation \eqref{T_eq} gives
\beq
\frac{M^2_{\rm 4D}}{4\pi T} = 2\Delta_{j,m} -1,
\label{M_4D}
\eeq
where the conformal dimension $\Delta_{j,m}$ is given by \eqref{dimV}. This coincides with  the result in \eqref{4D_masses}
obtained by world sheet methods, see \cite{SYlittles} for details.

Note that the Liouville interaction \eqref{Liouville_sup} is not taken into account in the tachyon action \eqref{tachyon_action}.
Therefore, we cannot find the spectrum of allowed values of $j$ ( see \eqref{discrete} and \eqref{principal}) from the equation \eqref{T_eq}. This spectrum is determined by the reflection from the Liouville wall.

Note also that the result \eqref{M_4D} together with the expression   \eqref{dimV} for the conformal dimension of $T_{j,m}$ is exact. To see this  
consider the region of large $s$, $s\to\infty$ in the equation \eqref{T_eq} where $g\to 1$. In this region we can   look for the solution for $T_{j,m}$ in the form \eqref{T_0} and repeat  the same derivation as above to get \eqref{M_4D}. Of course, at finite $\phi$, where $g$ becomes non-trivial, the expression for $T_{j,m}$ gets modified.
Moreover, we expect that the spectrum of allowed values of $j$ is also modified in the mass-deformed theory and depends on 
$(\phi_{\rm wall} -\phi_0)$.

\subsection{ Massless $b$-baryon}
\label{sec:b_deformed}

In this subsection we show that the massless in 4D $b$-baryon associated with the complex structure modulus of the conifold
survives the mass deformation. In terms of the Liouville theory it corresponds to the tachyon $T_{j,m}$ with $j=-\frac12$ $m=\pm \frac12$. Let us consider the case $m=-\frac12$ for definiteness. The $J_{-}=0$ for this state and it is described by a holomorphic function of $s$ at least in the limit $M\to\infty$. In other words it is a chiral primary field. 

Let us extrapolate this property to arbitrary $M$ and look for the holomorphic solution $T_b(s)\equiv T_{j=m=-1/2}(s)$. Equation \eqref{T_eq} reads in this case
\beq
T_b - \frac4{g}\,\pt_{\bar{s}}\Phi \,\pt_s T_b =0,
\label{b_eq}
\eeq
where we put the 4D mass $M_{\rm 4D}$ to zero. Calculating $\pt_{\bar{s}}\Phi$ using \eqref{Phi} and \eqref{pt_g} we get
\beq
\pt_{\bar{s}}\Phi =-\frac{Q}{4}\,g.
\label{pt_Phi}
\eeq
Then equation  \eqref{b_eq} takes the simple form
\beq
T_b+Q\,\pt_s T_b =0.
\label{T_b_eq}
\eeq
Observe now that the metric warp factor disappeared from the equation \eqref{T_b_eq}. Thus, its solution is the same as in the undeformed theory. Namely, we have 
\beq
T_b= e^{-\frac{\phi+iY}{Q}}.
\label{T_b}
\eeq
It coincides with the vertex $T_{j,m}$ in \eqref{T_0} with $j=m=-\frac12$ for $Q=\sqrt{2}$. The case of $m=\frac12$ corresponds to the complex conjugate of $T_b$.

Let us check the normalization of the $b$-baryon state over the Liouville dimension. Calculating its wave function \eqref{Psi} we get
for $Q=\sqrt{2}$
\beq
\Psi_b=e^{-\Phi}\,e^{-\frac{\phi+iY}{Q}}= \frac1{\sqrt{g}}\,e^{-\frac{iY}{Q}}, \qquad |\Psi_b|^2= \frac1{g},
\label{Psi_b}
\eeq
where we used \eqref{Phi}.
Then its norm is
\beq
\int d\phi \, dY\, g |\Psi|^2 =2\pi Q \int d\phi ,
\label{b_norm}
\eeq
where  factor $g$ arises due to the square root of the determinant of the metric.

Thus,  this state  is on the borderline between normalizable and non-normalizable states much in the same way as in the undefomed theory,
see Sec. \ref{sec:vertices}. In terms of the conifold radial coordinate this corresponds to the  logarithmically normalized state, see \eqref{Sb} and \cite{SYlittles} for details. 

We see that massless 4D baryon survives the mass deformation and is present in the 4D SQCD  at all values of mass $M$.
The (dressed) tachyon operator $T_{j=-\frac12, m=\pm \frac12}$ describes two scalar components of the BPS hypermultiplet in the 
4D \ntwo SQCD 
\cite{SYlittmult}. This leads us to the conclusion that the transition between regions of large and small $M$  is a smooth crossover rather then a sharp phase transition. The BPS state is  not affected (so analyticity is preserved), while the spectrum of non-BPS states associated with $T_{j,m}\neq T_b$ is expected to change as a function of   $(\phi_{\rm wall}-\phi_0)$.

Another related property of the solution  \eqref{T_b} is that the conformal dimension of the  operator in \eqref{T_b} is equal 
to 1/2 and therefore, 
as we mentioned in the end of Sec. \ref{sec:grav_sol}, the Liouville superpotential  \eqref{Liouville_sup} is not modified and remains  a marginal deformation of the mass-deformed Liouville theory, see \eqref{deformed_Liouville}.

\subsection{Schr\"odinger equation}

In this section we rewrite the tachyon equation \eqref{T_eq} in the form of the Schr\"odinger equation.
Substituting \eqref{pt_Phi} into  \eqref{T_eq} we get
\beq
\frac4{g}\,\pt_s\pt_{\bar{s}} T_{j,m} + Q( \pt_{\bar{s}} T_{j,m} +\pt_s T_{j,m})
+ 2\Delta_{j,m}\,T_{j,m} =0
\eeq
To get rid of terms with first derivatives we write
\beq
T_{j,m} = e^{\Phi}\, \tilde{\Psi}_{j,m},
\eeq
where the dilaton $\Phi$ is given by \eqref{Phi}.
With this substitution the equation for  $\tilde{\Psi}$ reads
\beq
-\left(\pt_{\phi}^2 + \pt_{Y}^2\right) \tilde{\Psi}_{j,m}+ Q^2\, g\left[(j+\frac12)^2 -m^2 -\frac14 (g-1)\right]
\tilde{\Psi}_{j,m} =0.
\label{tilde_Psi_eq}
\eeq

Since the warp factor $g$ does not depend  on $Y$ we can look for solutions of \eqref{tilde_Psi_eq} using the ansatz 
\beq
\tilde{\Psi}_{j,m} (\phi, Y) = e^{iQmY}\Psi_{j,m} (\phi).
\eeq
This gives the Schr\"odinger equation for the wave function $\Psi_{j,m}(\phi)$,
\beq
-\pt_{\phi}^2 \Psi_{j,m} + V_{{\rm eff}} (\phi)\Psi_{j,m} = E_{j} \Psi_{j,m},
\label{Psi_eq}
\eeq
 where the potential is given by
\beq
V_{{\rm eff}} (\phi) = -Q^2(g-1) \left[ m^2 -\left(j+\frac12\right)^2 +\frac{g}{4}\right],
\label{pot}
\eeq
while energy levels are determined by $j$,
\beq
E_j =-Q^2\left(j+\frac12\right)^2.
\label{E}
\eeq

As we already mentioned in Sec. \ref{sec:tachyon_eq} the Liouville interaction \eqref{Liouville_sup} is not taken into account
in the tachyon action \eqref{tachyon_action}, therefore we cannot determine energy levels and the spectrum of allowed $j$ from this equation in the region of large $M$,  where the Liouville interaction is essential. Instead we can use it in the region of small $M$  at $\phi_{\rm wall}\ll \phi_0$ or $M^2\ll 1/|b|^2$. In this region the Liouville interaction can be neglected and the energy levels and the spectrum of allowed $j$ can be found solving the equation \eqref{Psi_eq}. This will
give the  4D mass spectrum via Eq. \eqref{M_4D} which we interpret as a spectrum of hadrons in \ntwo SQCD with gauge group
U(4) and $N_f=8$.

The potential \eqref{pot} is attractive for $[m^2 -(j+\frac12)^2 + \frac14] >0$ and tends to zero at $\phi\to\infty$. 
Therefore, one may expect the  continues spectrum with $j=-\frac12 + is$ (see \eqref{principal}) with positive $E_j$ and the  discrete spectrum with negative $E_j$ \footnote{Note that we are looking for the spectrum of normalizable and borderline-normalizable states.}. 

However, the problem turns out to be  more complicated because near the singularity at $\phi\to\phi_0$ the potential \eqref{pot} is of the  Calogero type \cite{Calogero}  with the ''falling to the center'' behavior,
\beq
V_{{\rm eff}} (\phi)|_{\phi\to\phi_0} \approx  \frac{\alpha}{(\phi-\phi_0)^2}, \qquad \alpha= - \frac14 ,
\label{Calogero_pot}
\eeq
where we used \eqref{pole}.

The Hamiltonian with this potential has the scale invariance and therefore seems to have no discrete spectrum. 
However, the accurate definition of what is the self-adjoint  Hamiltonian leads to a well defined setup of the Calogero problem \cite{GitTyutVoron}. It turns out that the spectrum crucially  depends on the coefficient $\alpha$ in front of 
$1/(\phi-\phi_0)^2$. For example ''falling to the center'' occurs at $\alpha < - \frac14$ when the discrete spectrum is not bounded from below.
The coefficient  $\alpha =-\frac14$ (see \eqref{Calogero_pot}) represents a very special case. 
In this case there is only one discrete level \cite{GitTyutVoron}.

Thus, we expect that our Schr\"odinger equation \eqref{Psi_eq} has exactly one discrete level for each value of $m$ allowed by the representation \eqref{m} and Gliozzi-Scherk-Olive projection. The detail study of the string spectrum associated with this Calogero problem is left for a future work.

\section {Conclusions}
\label{sec:conclusion}
\setcounter{equation}{0}

In this paper we considered the  mass deformation of the string
theory for the critical non-Abelian vortex supported in
\ntwo SQCD with gauge group U(2) and $N_f = 4$ quark flavors. Our mass deformation 
interpolates in four dimensions between the above mentioned theory and \ntwo SQCD with gauge group U(4) and $N_f = 8$ quark flavors. Building on previous results that the Coulomb branch of the world sheet theory for the critical non-Abelian string in \ntwo SQCD with gauge group U(2) and $N_f = 4$  flavors is described by \ntwo Liouville theory we switch on the quark mass parameter $M$ and study the mass deformation of the Liouville theory, which boils down to the  $M$-dependent metric of its target space and the  $M$-dependent dilaton. 

To find  the mass-deformed metric and the dilaton for the true string vacuum we solve the effective supergravity equations of motion. 
The solution shows the presence of the naked singularity of the metric. Nevertheless, we show that the massless $b$-baryon
associated with the deformation of the complex structure of the conifold does not ''feel'' the metric deformation and remains massless
in the mass-deformed theory.

Next we present the Schr\"odinger equation for tachyon vertex operators which at $j\le -\frac12$ describes normalizable and borderline-normalizable string states. These
states correspond to hadrons living in 4D \ntwo SQCD. We give a qualitative discussion of the structure of the mass spectrum 
of tachyon states. In particular, we show that in the region of small $M$  finding the string spectrum is linked to the Calogero problem.

As a directions of future research we can mention the detail study of the string spectrum and its dependence on the  mass parameter $M$. 
 In particular, in the limit of small $M$ this spectrum gives the mass spectrum of hadrons in 4D \ntwo SQCD with gauge group U(4) and $N_f=8$ flavors of quarks.

Another challenging problem is to understand the physical nature of the naked singularity of the Liouville target space metric and its possible resolution.

\section*{Acknowledgments}

The author is grateful to  E. Ievlev and  A. Marshakov  for very useful and 
stimulating discussions. This work  was  supported  by the Foundation for the Advancement of Theoretical Physics and Mathematics ''BASIS'',  Grant No. 22-1-1-16-1.

%
%




\renewcommand{\theequation}{A.\arabic{equation}}
\setcounter{equation}{0}

\end{document}